\documentstyle[12pt,psfig]{article}
\begin{document}

\begin{center}
\Large\bf
CHIRAL SYMMETRY AND THE NUCLEON-NUCLEON 
INTERACTION\footnote{Invited
Talk presented at the {\sc XIII International Seminar
on High Energy Physics Problems} at the {\it Joint Institute
for Nuclear Research}, Dubna (Russia), September 2-7, 1996.}
\\
\vspace*{.5cm}
{\large\sc R. Machleidt}\footnote{e-mail address: machleid@phys.uidaho.edu}
\\
\vspace*{.2cm}
{\large\it Department of Physics, University of Idaho,}\\
{\large\it Moscow, Idaho 83843, U. S. A.}
\end{center}

\begin{abstract}
The main progress in the field of nucleon-nucleon (NN)
potentials, which we have seen in recent years,
is the construction of some very quantitative
(high-quality/high-precision) NN potentials.
These potentials will serve as excellent input for
microscopic nuclear structure calculations and will allow for
a systematic investigation of off-shell effects.
After this enormous quantitative work, it is now time
to re-think the NN problem in fundamental terms. 
We need a derivation of the nuclear force which observes
Lorentz invariance and the symmetries of QCD.
\end{abstract}

\subsection*{Current Status of NN Potentials}

\paragraph*{The New High-Precision NN Potentials.}
In 1993, the Nijmegen group published a phase-shift analysis
of all proton-proton and neutron-proton data below
350 MeV lab.\ energy with a $\chi^2$ per datum of 0.99
for 4301 data~\cite{Sto93}.
Based upon this analysis, charge-dependent NN potentials have
been constructed by the Nijmegen~\cite{Sto94}, the
Argonne~\cite{WSS95}, and the Bonn (CD-Bonn~\cite{MSS96})
groups which reproduce the NN data with
a $\chi^2$/datum $\approx 1$ (see Table~1).
The main difference between these new potentials is that some
are local and some non-local.

Ever since NN potentials have been developed,
local potentials have enjoyed great popularity
because they are easy to apply in configuration-space
calculations.
Note, however, that numerical ease is not a proof for the local nature
of the nuclear force.
In fact, any deeper insight into the reaction mechanisms underlying
the nuclear force suggests a non-local character.
In particular, the composite structure of hadrons should lead
to large non-localities at short range.
But, even the conventional and well-established
meson theory of nuclear forces---when derived properly 
relativistically and without
crude approximations---creates a non-local interaction.

\begin{table}[t]
\caption{Recent high-precision NN potentials and predictions for
the two- and three-nucleon system.}
\small
\begin{tabular}{lccccc}
\hline\hline
              & CD-Bonn[4]& Nijm-II[2]& Reid'93[2]& $V_{18}$[3] & Nature\\
\hline
\hline
Character  & non-local   & local  & local          & local &non-local \\
$\chi^2$/datum  & 1.03 & 1.03     & 1.03           & 1.09  & --   \\
$g^2_\pi/4\pi$  & 13.6 & 13.6     & 13.6           & 13.6  & 13.75(25) \\
\hline
{\it Deuteron properties:} \\
Quadr.\ moment (fm$^2$) & 0.270 & 0.271 & 0.270 & 0.270  & 0.276(3)$^a$ \\
Asymptotic D/S state & 0.0255  & 0.0252 & 0.0251 & 0.0250 & 0.0256(4) \\
D-state probab.\ (\%) & 4.83& 5.64 & 5.70  & 5.76     &  --    \\
\hline
{\it Triton binding (MeV):} \\
non-rel.\ calculation & 8.00   & 7.62  & 7.63 & 7.62   & --     \\
relativ.\ calculation & 8.19 & -- & -- & -- & 8.48\\
\hline\hline
\multicolumn{6}{l}{\footnotesize
$^a$ Corrected for meson-exchange currents and relativity.}
\end{tabular}
\normalsize
\end{table}

In Fig.~1, we show the half off-shell
$^3S_1$--$^3D_1$
potential that can be produced only by tensor forces.
The on-shell momentum $q'$ is held fixed at 153 MeV
(equivalent to 50 MeV lab.\ energy),
while the off-shell momentum $q$ runs from zero
to 1400 MeV.
The on-shell point ($q=153$ MeV) is marked by a solid dot.
The solid curve is the new relativistic one-boson-exchange (OBE) 
potential,
CD-Bonn~\cite{MSS96}. When the relativistic
one-pion-exchange
(OPE) amplitude is replaced by the static/local
approximation
the dashed curve is obtained.
When this approximation is also used for the one-$\rho$
exchange, the dotted curve results.
It is clearly seen that the static/local approximation
substantially increases the tensor force off-shell.
Obviously, relativity and non-locality
are intimately interwoven.

In Table I (upper part), we summarize some important
two-nucleon properties as predicted
by the new high-quality potentials.
Using the same $\pi NN$ coupling constant,
all potentials predict almost identical deuteron observables 
(quadrupole moment
and asymptotic D/S state normalization). Note, however,
that the (un-observable) deuteron
D-state probability comes out significantly larger  
for the local potentials ($\approx 5.7$\%)
as compared to
the non-local CD-Bonn potential (4.8\%).
Obviously, the deuteron D-state probability
is kind of a numerical measure for
the off-shell strength of the tensor force,
shown graphically in Fig.~1.

\begin{figure}[t]
\hspace*{3.5cm}\psfig{file=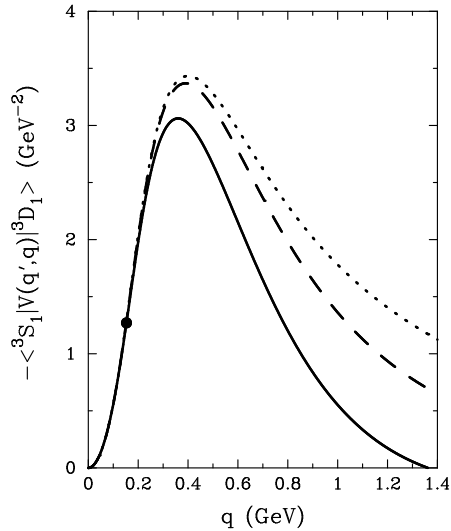,height=6cm}
\caption{Half off-shell $^3S_1$--$^3D_1$ amplitude
for the relativistic CD-Bonn potential (solid line).
The dashed curve is obtained when the local
approximation is used for the one-pion-exchange (OPE) and the dotted curve
when OPE and one-$\rho$
exchange are both local. $q'=153$ MeV.}
\end{figure}

\paragraph*{Off-Shell NN Potential and Nuclear Structure.}

By construction, NN interactions reproduce the two-nucleon
scattering data and the properties of the deuteron.
Assuming the existence of NN scattering data of increasing
quantity and quality, the NN interaction can be fixed with arbitrary
accuracy---``on the energy shell'' (on-shell). 
However, in nuclear structure the NN potential contributes also
off-shell.
For several decades, the qustion has been around, how large
the effects can be that come from differences in the off-shell
behavior of different NN potentials when applied in
microscopic nuclear structure calculations.
The recent construction of high-precision NN potentials
has finally set the stage for a reliable investigation
of the issue.

Friar {\it et al.}~\cite{Fri93} have calculated the binding
energy of the triton (in charge-dependent 34-channel
Faddeev calculations)
applying the new, high-quality Nijmegen and Argonne
potentials and obtained
almost identical results for the various local models,
namely,
$7.62\pm 0.01$ MeV (experimental value: 8.48 MeV), 
where the uncertainty of $\pm 0.01$ MeV
is the variation of the predictions which occurs when 
different local potentials are used.
The smallness of the variation is due to the fact that all the local
potentials 
have essentially the same off-shell behavior.

Using the new, non-local CD-Bonn potential,
we have performed a (34-channel, charge-dependent)
Faddeev calculation for the triton 
and obtain 8.00 MeV binding energy (cf.\ Table I).
This is 0.38 MeV more than local potentials
predict. 
The unaquainted observer may be tempted to believe
that this difference of 0.38 MeV is quite small,
almost negligible. However, this is not true.
The difference between the predictions by local potentials
(7.62 MeV) and experiment (8.48 MeV)
is 0.86 MeV.
Thus, the problem with the triton binding is that
0.86 MeV cannot be explained in the simplest way,
that is all.
Therefore, any non-trivial contribution
must be measured against the 0.86 MeV gap between experiment
and simplest theory.
On this scale, the non-locality considered in this investigation
explains 44\% of the gap; i.~e., it is substantial 
with respect to the discrepancy.

The above three-body results were obtained by using the
conventional non-relativistic Faddeev equations.
However, since CD-Bonn is a relativistic potential,
one can also perform
a relativistic Faddeev calculation by extending
the relativistic three-dimensional Blanckenbecler-Sugar
formalism to the three-body system~\cite{SXM92}. 
The binding energy prediction by CD-Bonn
then goes up to 8.19 MeV. 
This further increase can be understood as an additional off-shell effect
from the relativistic two-nucleon $t$-matrix applied in the three-nucleon
system.

The trend of the non-local Bonn potential to increase binding
energies has also a favorable impact on predictions for
nuclear matter~\cite{Mac89} and ground/excited states of
nuclei~\cite{Jia92}.

\subsection*{Chiral Symmetry and the Nuclear Force}

\paragraph*{Introduction.}
The nuclear force is a sample of strong interactions.
The fundamental theory of strong interactions is QCD.
Thus, ideally, the NN interaction should be derived directly
from QCD.

The last decade has seen numerous attempts to derive
the nuclear force from QCD or
``QCD-inspired'' models---with mixed success.
The suggestion has then been made, to take an alternative attitude,
namely: If we can't solve QCD, then let's at least observe its symmetries,
particularly, chiral symmetry. The most outstanding advocate of this
view is S. Weinberg~\cite{Wei79}.

Past models for the NN interaction do not have any
clear relationship with chiral symmetry.
Since all NN models fit the data,
they are also not in gross violation of chiral symmetry.
The best example for this `accidental' compliance with
chiral symmetry 
is the fact that essentially all field-theoretic
models for the NN interaction leave out pair terms---because
they don't fit.

A recent fundamental progress in nuclear physics is that chiral symmetry
is now taken seriously.
Therefore, it is about time to relate the NN interaction to chiral
symmetry in a way which is better than just accidental.

Even though a review article entitled
``Chiral Symmetry and the Nucleon-Nucleon Interaction''
was published already some 17 years ago~\cite{Bro79},
there have been few serious attempts to pursue this line
of research~\cite{DBS84}.

Very recently, by initiative of
Steven Weinberg~\cite{Wei90}, a group at the University of Texas
\cite{ORK94} has
derived a NN potential
using chiral perturbation
theory. 
In this work, the properties of the deuteron
and the NN phase shifts below 300 MeV lab.\ energy are reproduced
qualitatively using 26 parameters.
The significance of this work is that it represents the first comprehensive
attempt to strictly base the NN interaction on chiral symmetry.
Even though this attempt can been deemed as basically successful,
there are reasons for concern.

Conventional field-theoretic models for the NN interaction
have typically 12 parameters and yield an accurate fit of the 
deuteron and NN scattering up to about 300 MeV.
The Texas potential~\cite{ORK94} has twice as many parameters
and, nevertheless, provides only a qualitative fit.
The qualitative nature of the fit makes this potential unsuitable 
for nuclear structure
applications.

Of even more concern is another point:
the Texas model~\cite{ORK94} is nonrelativistic.
We have seen in our above discussion on off-shell 
effects in nuclear structure that
off-shell momenta
up to about 2 GeV are important, making a relativistic
approach mandatory.
Moreover, the proper explanation of
nuclear saturation requires a relativistic
approach~\cite{Mac89}.

\paragraph*{What is an Appropriate Lagrangian?}
Thus, we need a Lagrangian which fulfils the following minimal
requirements (besides the usual ones, like, parity conservation
and inclusion of nucleons):
\begin{itemize}
\item
Lorentz invariance,
\item
(approximate) chiral symmetry,
\item
inclusion of (heavy) vector bosons.
\end{itemize}
Note that this `minimal' program in regard to what is needed for
a realistic nuclear interaction is, in fact, a maximal program
in terms of chiral symmetry.
First work on chiral symmetry in nuclear physics
typically considered only pions
and interactions between them. In the next step, also baryons
were included, which was already a big and difficult step.
Now, we even want to include (heavy) vector bosons.
So far there is little work in this direction;
which makes it a real challenge.
Note that the inclusion of heavy bosons is crucial
for a realistic description of the nculear force.
One reason why
the Texas potential is not doing too well in quantitative terms
inspite of its large
number of parameters may be the omission
of vector bosons.

Recently, Furnstahl, Serot, and Tang~\cite{FTS96} have
proposed the following Langrangian which fulfils all of
the above requirements:

\begin{eqnarray}
{\cal L} & = & 
 \bar{N} (i\gamma^\mu {\cal D}_\mu 
 + g_A \gamma^\mu \gamma_5 a_\mu
 - g_\omega \gamma^\mu \omega_\mu
 - M ) N  
 \nonumber \\
 & &
 - \frac{f_\rho}{4M} \bar{N} \rho_{\mu\nu}\sigma^{\mu\nu} N
 - \frac{f_\omega}{4M} \bar{N} \omega_{\mu\nu}\sigma^{\mu\nu} N
 \nonumber \\
 & &
 + \frac{F^2_\pi}{4} tr (\partial_\mu U \partial^\mu U^\dagger)
 - \frac14 \omega_{\mu\nu}\omega^{\mu\nu}
 + \frac12 m_\omega^2 \omega_{\mu}\omega^{\mu}
 \nonumber \\
 & &
 - \frac12 tr (\rho_{\mu\nu} \rho^{\mu\nu})
 + m_\rho^2 tr (\rho_{\mu} \rho^{\mu})
 - g_{\rho\pi\pi} \frac{2F^2_\pi}{m^2_\rho} tr (\rho_{\mu\nu}v^{\mu\nu})
 + \cdots
\end{eqnarray}

with

\begin{equation}
\begin{array}{lcllcl}

\pi & \equiv & \mbox{\boldmath $\tau \cdot \pi$} /2 
\hspace*{2.5cm} 
&
{\cal D}_\mu & = & \partial_\mu + iv_\mu 
 + ig_\rho \rho_\mu 

\\
   \xi & = & e^{i\pi/F_\pi}  
&
v_\mu & = & -\frac{i}{2} (
                           \xi^\dagger \partial_\mu \xi
                        +  \xi \partial_\mu \xi^\dagger)  

\\

U & = & \xi \xi 
&
a_\mu & = & -\frac{i}{2} (
                           \xi^\dagger \partial_\mu \xi
                        -  \xi \partial_\mu \xi^\dagger)  

\\

   \rho_\mu & \equiv & \mbox{\boldmath $\tau \cdot \rho$}_\mu /2  
&
v_{\mu\nu} & = & \partial_\mu v_\nu - \partial_\nu v_\mu
                    + i [v_\mu, v_\nu] 

\\

\sigma_{\mu\nu} & \equiv & \frac{i}{2}[\gamma_\mu, \gamma_\nu]  
&
\tilde{D}_\mu \rho_\nu & \equiv & \partial_\mu \rho_\nu 
                                  + i [v_\mu,\rho_\nu]  

\\

\omega_{\mu\nu} & = & \partial_\mu \omega_\nu - \partial_\nu \omega_\mu  
&
\rho_{\mu\nu} & = & \tilde{D}_\mu \rho_\nu - \tilde{D}_\nu \rho_\mu  
                                  + i g[\rho_\mu,\rho_\nu]  

\end{array}
\end{equation}
where $N=\left( \begin{array}{c} p \\ n \end{array} \right)$
is the nucleon field with $p$ and $n$ the proton and neutron
fields, respectively; {\boldmath $ \pi $} are the Goldstone pion
fields which form an isovector, and {\boldmath $ \rho $} are
the rho-meson fields.
$\omega_\mu$ is the field of the omega meson which can be considered
as a chiral singlet in chiral SU(2) symmetry.
$F_\pi \approx 93$ MeV is the pion decay constant,
$g_A \approx 1.26$ is the axial coupling constant, and
$M$ denotes the nucleon mass.
$g_\rho$ is the vector-coupling constant of the
$\rho$ meson to the nucleon and $f_\rho$ is the tensor-coupling
constant; similarly for the $\omega$.
$m_\rho$ and $m_\omega$ are the masses of the $\rho$ and
$\omega$ mesons, respectively.
$g_{\rho\pi\pi}$ is the $\rho\pi\pi$ coupling constant.

The Lagrangian Eq.~(1) is essentially
based upon the classic papers by Weinberg~\cite{Wei68}
as well as Callan, Coleman, Wess, and Zumino~\cite{Cal69}.

The ellipsis in Eq.~(1)
stands for higher-order terms involving additional powers of the vector
fields and their derivatives, and other terms consistent with
chiral symmetry.
For example, so-called
`contact' terms of the kind
$\bar{N}N\bar{N}N$ and 
$\bar{N}\gamma_\mu N\bar{N}\gamma^\mu N$ 
are permitted by chiral symmetry.
However, these terms could be absorbed by the exchange of a 
scalar-isoscalar boson and the omega meson, respectively, which may be
justified by the success of the one-boson-exchange model
for the NN interaction.

For the intermediate range attraction of the nuclear force,
there are two suitable scenarios:
it can be constructed from correlated two-pion exchange
(as implied by the pion terms in the above Langrangian)~\cite{CPS92};
or a scalar-isoscalar field 
(to approximate
the correlated two-pion exchange)
can be introduced 
in such a way as to maintain chiral symmetry;
note that this scalar field is not the chiral partner of the pion.

When applying the Langrangian Eq.~(1) to the NN problem,
suitable expansion parameters
have to be identified.
This could be patterned after chiral perturbation theory.
Naive dimensional analysis and the assumption of ``naturalness''
for the coupling constants could serve as guidelines.
However, since this is a relativistic approach, there are
characteristic differences to chiral perturbation theory
and, therefore, the details still have to be worked out.
It is suggestive to consider an expasion in powers of the
fields and their derivatives.

In lowest order, the Lagrangian Eq.~(1) will recover the terms
familiar from the traditional one-boson-exchange model
(however, with the pion coupled to the nucleon via gradient coupling).
Higher orders will generate new terms characteristic for
the chiral approach. These terms have to be identified
systematically
and, then, added to the two-nucleon scattering and bound state problem.
In this way, a relativistic NN model consistent with chiral symmetry
will be developed step by step,
and checked against the experimental NN data.

\begin{figure}
\hspace*{3.5cm}\psfig{file=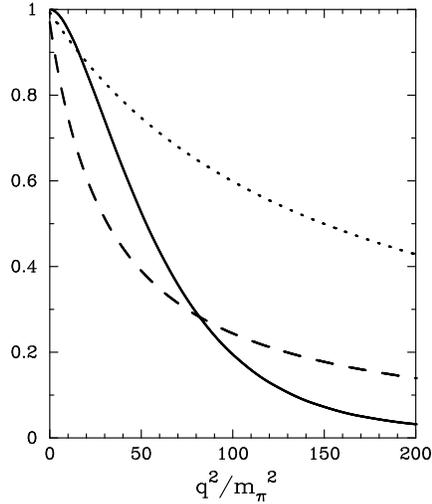,height=6cm}
\caption{Comparison of different $\pi NN$ form factors.
The solid line represents the form factor extracted from the
Skyrme model. The dashed and dotted lines are conventional
monopole form factors with cutoff masses $\Lambda = 0.8$
and 1.7 GeV, respectively.}
\end{figure}

\paragraph*{The $\pi NN$ Form Factor.}

Analyzing the meson-baryon scattering $S$-matrix in the soliton
sectors of effective, chiral Lagrangians does not require to 
consider separately
 meson-baryon form factors (FFs) because the soliton
solution takes care of the spatial structure of the interaction in a 
consistent way~\cite{HPJ90}. This holds, 
of course, also for the analysis of the baryon-baryon interaction.
Thus, one can extract 
meson-baryon FFs from 
soliton solutions of mesonic actions, which would allow for a comparison 
with FFs
typically used in conventional meson-exchange models of the baryon-baryon 
interaction.
Early work on this topic yielded
 rather disappointing results:
The 
values for the cutoff mass $\Lambda$ of an equivalent monopole cutoff
came out to be around 0.6--0.8 GeV which is
less than half of the 1.3--1.7 
GeV typically required in OBE models for the NN interaction~\cite{Mac89}.

However,
inspection of $\pi N $ P-wave scattering~\cite{HPJ90} shows that the 
FF which actually 
determines the dominant part of the scattering amplitude
contains an additional {\it metric factor} which has a profound influence on 
its shape, such that it 
is no longer compatible with the monopole form. 
As demonstrated in Fig.~2,
the FF extracted from the Skyrme model (solid line)
has a very small slope near
$q^2=0$ and the curvature is negative.
This means that for small $q^2$ the effective $\pi NN$ coupling strength
stays much closer to its value at $q^2=-m_\pi^2$ than for comparable monopole
FFs. 

We have investigated
to which extent such FF could be helpful in 
conventional meson-exchange potentials
for nucleon-nucleon scattering and the structure of nuclei.
It is well known that
soft ($\Lambda \approx 0.8$ GeV) monopole FFs fail
in the NN system, since they cut out too much of the tensor force
provided by the pion: the deuteron quadrupole moment and asymptotic
D/S state ratio and the $\epsilon_1$ mixing parameter of NN scattering
(which all depend crucially on the nuclear tensor force) come out too 
small~\cite{Mac89}. 
First results indicate that
the very hard behaviour of the Skyrme model FF for small $q^2$ 
is extremely useful in the OBE model for the
NN interaction.
On the other hand the very soft behaviour for 
$q^2 >$ 50 $m_\pi^2$ cuts off higher momenta much 
more efficiently than typical hard monopole FFs (cf.\ Fig.~2).

A consequence of this is that
the Skyrme FF allows to describe the $\pi N$ and $NN$ system
quantitatively with the {\it same} FF. 
Note that this is not possible with
soft monopole FF unless 
a heavy pion, $\pi'(1200)$ (representing $\pi-\rho$ correlations),
is introduced.

\subsection*{Summary}

Several high-quality/high-precision NN potentials are now available
which fit the low-energy NN data with idential perfection.
These potentials differ, however, in their off-shell
behavior. Thus, the stage is set for a systematic investigation
of off-shell effects in microscopic nuclear structure.

After the excessive quantitative work on `perfect' NN potentials
which we have seen during the past five years,
it is now time to sit back and {\it think} again---about
the more fundamental aspects concerning the nuclear force.
We need a derivation of the nuclear force which is Lorentz
invariant and consistent with the symmetries of QCD.

This work was supported in part by the NSF (PHY-9211607).

\end{document}